\documentstyle[12pt]{article}
\begin{document}
\begin{center}
{\bf TWO-PHOTON PROCESSES IN TWO-NUCLEON SYSTEMS AS A TOOL OF SEARCHING
FOR EXOTIC SIX-QUARK RESONANCES }
\footnote{This work was supported by
the RFBR grants No. 96-02-19147 and 96-15-96423.}
\end{center}
\begin{center}
\vspace*{5mm} S.B. GERASIMOV\\
{\it Bogoliubov Laboratory of Theoretical Physics,\\}
{\it Joint Institute for Nuclear Research, Dubna, 141980 Russia}
\end{center}
\vspace*{.5mm}
\begin{abstract}
The reaction $pp \to pp2\gamma$, proposed earlier to probe for
the NN--decoupled dibaryon resonances, has been studied by the DIB$2\gamma$
Collaboration (JINR) and their preliminary data seem to give
evidence for the resonance effect at about 1920 MeV. We discuss
some other two-photon processes: the double radiative capture
reaction in pionic deuterium, photoabsorption sum rules and
the nuclear Compton scattering, which we consider to be feasible
for further test and invesigation of possible low-lying,
exotic six-quark states.
\end{abstract}


\section{Introduction}
The experimental discovery of dinucleon (or, generally, multibaryon)
resonances not decaying into two (or more) nucleons in the ground state
would be one of the most spectacular explications of nonpotential, nonnucleon
degrees of freedom and would imply important consequences
in further development of nuclear and hadron matter physics.
The nonstrange NN-decoupled dibaryons with small widths appear to be the most
promising and interesting candidates for experimental searches. Among
available candidates to be confirmed (or rejected) in future dedicated and,
hopefully, more sensitive experiments we wish to mention the indications
of the existence of a narrow $d^{\prime}$ dibaryon slightly above the $\pi NN$
threshold, coming from data on pion double charge exchange on nuclei
\cite{a1}, and $d_{1}^{\ast}$-enhancement below $\pi NN$-threshold, seen in
preliminary data on the proton-proton double bremsstrahlung
at $200 MeV$ \cite {a2}.\\
It would be of undoubted interest to try different reactions to search
for these states. With the pion
absorption reactions, the presumed $d_{1}^{\ast}$-dibaryon
can be excited via strong interactions only  on the three- (or more) nucleon
states. With the deuteron targets most thoroughly investigated and most
easy to deal with theoretically, we have to resort to radiative processes.

This report aims at drawing attention to the real feasibility and suitability
of study of radiation reactions such as $\pi (\mu)$-meson capture processes
in the deuterium mesoatoms and the photon-deuteron scattering or analysis
of integral characteristics of corresponding photoabsorption cross sections
for the sake of inquiry on possible nucleon and dinucleon exotics.

\section{The $2\gamma$-production in $pp$-interactions: indications from
data}

For the sake of completeness we remind some points referring to
the double photon emission reaction.
In refs.\cite{a3,a4} the process $pp \to \gamma {^2B} \to pp \gamma \gamma$
has been advocated to provide
unique possibilities of searching for and investigating narrow
NN-decoupled dibaryon resonances with masses below the pion production
threshold. The method of searching for these  narrow dibaryon
resonances is based on the measurement of the photon
energy distribution in the $pp \to pp \gamma \gamma$ reaction by detecting both
photons in coincidence. The narrow dibaryons, if they exist, should be seen
as sharp $\gamma$-lines against a smooth background due to
the photons from the radiative resonance decays($^2B \to \gamma pp$)
and double pp-bremsstrahlung with an anticipated good signal-to-background
ratio. The position of this line depends on the energy of the incident
proton and the resonance mass M$_B$. Its width is determined by the total
width of resonance $\Gamma_{tot}$ and energy resolutions of the experimental
setup.
The $\gamma$-ray energy spectrum for the
$pp \to pp \gamma \gamma$ reaction at the proton energy $\sim 200$ MeV has
been measured in ref.\cite{a2}. A distinct enhancement at the photon
energy about 42 MeV was observed in measured energy spectrum.
It can be interpreted as a signal due to the narrow exotic
dibaryon $^2B$ formation and decay in the $pp \to \gamma ^2B \to pp \gamma
\gamma$ processes. Distribution of the dibaryon mass obtained under this
assumption shows a narrow peak with mass $M_B$=1923.5$\pm$4.5 MeV and width
FWHM=31.3$\pm$5.0 MeV.
The statistical significance of this peak exceeds 8$\sigma$.
In ref.\cite{a6}, the radiative transition
$d_{1}^{\ast}(IJ^{P}=11^{+}) \to \gamma pp$ or
the inverse reaction $pp \to \gamma d_{1}^{\ast}$
has been considered as a two-step process, where the
presumably lowest pp-decoupled state with the $J^P=1^{+}$, tentatively called
$d_{1}^{\ast}$, is coupled with the initial or final hadron states through the
intermediate $N \Delta$-state with the same quantum numbers.The $"\Delta"$ -
symbol may be referred also to the virtual $\pi N$-complex with quantum
numbers of the $\Delta(1232)$-resonance but a different invariant mass.
The $d_{1}^{\ast}\Delta N$-vertex is described by a simple form of the
quasi-two-body wave function, for which the Hulthen-type radial
dependence was chosen by analogy with the deuteron radial wave function:
\begin{equation}
R(r) = N\frac{1}{r}exp(-\alpha r)(1 - exp(-\beta(r-r_c)))
\end{equation}\label{Hult}
where N is the normalization constant,$\alpha=\sqrt{2M_{red}\varepsilon_{1}},
\varepsilon_{1}=M+M_{\Delta}-M_{d_{1}^{\ast}},
M_{red}^{-1}=M^{-1}+M_{\Delta}^{-1},\beta=5.4 fm^{-1}, r_{c}=.5 fm$ and
$R(r)=0$ for $r \leq r_{c}$ is understood. The second factor in Eq.(2),
representing the behavior of wave function in the "interior" region
outside the hard core with the radius of $r_c=.5 fm$ is taken quite similar
to the deuteron case.Taking $M_{d_{1}^{\ast}} \simeq 1920 $~MeV for granted,
the transition magnetic moment $\mu(p \Delta^{+})=2\sqrt{2}/3 \mu(p)$
according to the $SU(6)$-symmetry and plane waves for initial protons,
we get an estimation
\begin{equation}
(\frac {d\sigma}{d\Omega_{\gamma}})_{c.m.} \simeq \frac
{\alpha(W-M_{d_{1}^{\ast}})^3(\mu(p)I(q))^2}{9Mq} \simeq 40 \frac{nb}{sr}
\end{equation}
$$ I(q) = \int_{r_c}^{\infty}dr r^2R(r)\frac{\sin(qr)}{qr}$$
where $\alpha=1/137, \mu(p)=2.79,q \simeq \sqrt{MT_{lab}/2}$,
$M$ is the mass of the proton.
It turns out justified to neglect the retardation corrections,
i.e. we are using the long-wave approximation for the matrix element of
magnetic-dipole transition.Further,the result does not depend strongly on
variation of the "effective" mass $M_{\Delta}$ from $M+m_{\pi}$ to
$M_{\Delta}=1232 MeV$.With the integrated luminosity $L=10^{37} cm^{-2}$,
the data gave about 130 events after integration over energies of both
photons in the interval  $10 \leq \omega_{i} \leq 100 MeV$.Assuming the
spherical-symmetric distribution of photons, we obtain an estimate
of the cross section for one of photons to be registered
in the element of solid angle
\begin{equation} \frac
{d\sigma}{d\Omega_{\gamma}} \simeq 4\pi(\frac {d\sigma}{d\Omega_{\gamma_1}
d\Omega_{\gamma_2}})_{exp} \simeq 65\frac{nb}{sr}
\end{equation}
By definition, this is the quantity which should be confronted
with Eq.(2) and we estimate the result of this comparison as a reasonable
one.

Concerning the negative result of the recent experiment at CELSIUS \cite{Ce},
we note that this experiment was not a dedicated one for a search of the
two-photon production in proton-proton interactions, but, rather, it was
 optimised for investigation of the $pp \to pp\gamma$ reaction. Therefore,
in our opinion, the immediate relation of their upper bound on the exotic
dibaryon production cross section with the result of ref.\cite{a2}
still needs to be developed.

\section{Meson capture from mesoatomic states and possible exotics}

In what follows we concentrate mainly on the radiative pion capture
processes because they are experimentally easier to investigate.
This proposal is not entirely new. In fact, there is the work devoted
to search for the exotic isotensor ( with the isospin $I=2$) dibaryon
resonance done at TRIUMF \cite{a5}. We, however, propose to look for a rather
specific object, which seemed to be beyond the design and kinematics area
covered by the abovementioned experiment. Furthermore we relay on what
can be taken into account from an analysis \cite{a6} of situation
connected with the double bremsstrahlung experiment \cite{a2}.\\
A few remarks are to be made about quantum numbers of the mentioned
candidates.  The isospin 0 assignment for the $d'$ resonance with quantum
numbers $J^P=0^{-}$ was motivated by calculations within a QCD string
model, while the isospin $2$ assignment was made in \cite{a7} on the basis
of the $\pi NN$ bound system molel and within the Skyrme model approach.  With
the $d'$ quantum numbers $0^{-}$ and isospin 2, the dominant isobar in the
pion-nucleon sub-system is $P_{33}$ or $\Delta$, so that nucleon and
isobar have relative orbital angular momentum 1 and total spin 1.  With the
isospin 0 assumed, the dominant pion-nucleon sub-system in three-body model is
the $S_{11}$ isobar and the nucleon-isobar quasi-two-body
system has the relative orbital momentum and total spin both equl to 0.\\
For
the $d_{1}^{\ast}$-state, presumably seen in the proton-proton double
bremsstrahlung reaction below the pion threshold, we suggest the $\Delta N$
with relative orbital angular momentum 0 as the dominant cluster
configuration.  Of the possible values 1 or 2  for spin and isospin, we
consider the unit spin/isospin value as more natural for the state with lowest
mass.  In this case the NN decay channel is strictly forbidden by the
exclusion principle, and if the $\pi NN$ decay mode is kinematically forbidden
we have the radiative decay as the only possible one with the width of $\sim$
KeV scale.  It might, however, be that the isospin 2 assighment for
$d_{1}^{\ast}$ is
dynamically preferable, like in the case of the $d'$-state mentioned above.
In that case the $d_{1}^{\ast} \to NN $ decay mode is possible though strongly
suppressed being of order $O(\alpha ^2)$ versus the radiative decay of the
order $O(\alpha)$. To discriminate between two options $(I=0$ vs $2)$, it is
important to study several reations differently sensitive to different isospin
values. It is with this feature in view, that we propose to make use of the
double radiative capture process  in the pionic deuterium in addition to the
nucleon-nucleon double bremsstrahlung reactions. As will be seen in the next
section, the transition operator between the deuteron and dibaryon resonance
state vectors is transformed, by construction within the assumed model, as the
isovector under rotations in the isospin space.  This means that, unlike the
proton-proton double bremsstrahlung reaction, the narrow dibaryon will not be
excited in the radiative pion capture on the isosinglet deuteron if the
isospin of this resonance equals 2.

The branching ratios of different
$\pi^{-}$ - capture channels in the pionic deuterium: $\pi^{-}d \to nn(.7375
\pm .0027), nn\gamma (.2606 \pm .0027), nne^{+}e^{-} ((1,81 \pm .02)\cdot
10^{-3}), nn\pi^{0}((1.45 \pm .09)\cdot 10^{-4})$, measured at TRIUMF
\cite{a8,a9}, give the total probability $w(\pi^{-}d \to X(measured) = 1.000
\pm .038$. Therefore, the upper bound of still undetected channels is not much
larger than $\simeq .38 \%$, well within capability of measurements if there
is good signature .  We estimate which could be the yield of two $\gamma's $
from consequent processes of the radiative excitation and de-excitation of the
exotic $d_{1}^{\ast}(1920)$, presumably seen in the proton-proton double
bremsstrahlung:
\begin{eqnarray} (\pi^{-}d)_{atom} \to \gamma d_{1}^{\ast} \to
\gamma \gamma X
\end{eqnarray}
The radiative decay branching ratio $Br(d_{1}^{\ast} \to
\gamma X)=1$, hence we need to calculate only the $d_{1}^{\ast}$ - excitation
probability, i.e. the transition $(\pi_{-}d)_{atom} \to \gamma
d_{1}^{\ast}(1920)$.
The radiated photon takes off the energy $\omega = 92.9 MeV$ thus enabling the
resonance state to be on its mass-shell.As a hint for possible qualitative
estimate we note that the probability of ordinary radiationless
$\pi^{-}$-capture $w((\pi^{-}d)_{atom} \to nn)$ is only three times as large
as that of the radiation capture.This is indication of the dominantly
short-range nucleon-nucleon interaction dynamics involved in a pure strong
capture channel, resulting in a poor overlap with the deuteron wave function
having characteristically large spatial extensions. An essential feature of
mechanisms of both the ordinary $NN$-channel \cite{a10} and
the assumed $d_{1}^{\ast}$-excitation is the appearance of the
$N \Delta$-configuration in an intermediate state of reactions considered.
So, it seems reasonable to expect that
\begin{eqnarray}
BR((\pi^{-}d)_{atom} \to \gamma d_{1}^{\ast}) \sim \alpha_{em}
BR((\pi^{-}d)_{atom} \to nn) \simeq .74/137 \simeq .5 \% .
\end{eqnarray}

More quantitative
estimation is made within a model used previously \cite{a6} for the reaction
$pp \to \gamma d_{1}^{\ast}$.  Namely, we assume the reaction mechanism when
$\pi^{-}$ is radiatively captured by a nucleon to form the (virtual) $\Delta$
that, in turn, is associated with a spectator nucleon to form the
$d_{1}^{\ast}(1920)$-resonance.

As in the case of our estimation of the $pp \to pp2\gamma$ cross section
in the previous section, we adopt the explicitly phenomenological approach
in calculation of the pion capture rate.

Having in mind the completeness of colourless hadron states, we
estimate the probability of the radiative transition $(\pi^{-}d)_{atom}
\to \gamma d_{1}^{\ast}(IJ^{P}=11^{+} $ as a two-step process,where the
presumably lowest $NN$-decoupled state with the $J^P=1^{+}$, we are calling
$d_{1}^{\ast}$, is coupled with the initial or final hadron states through the
intermediate $N \Delta$-state with the same quantum numbers.
We make use of standard formulas for a capture from atomic states following
from the assumption that the hadronic reaction is much shorter in range
than the atomic orbit radii. The rate is, schematically,
\begin{eqnarray}
w(L=0&\mbox{atomic state}) \sim |\psi(0)|^2 |\langle f|T_{\pi\gamma}(\vec{0})|i\rangle |^2
\end{eqnarray}

where $\psi(r)$ is the $L=0$ pionic atom wave function,
$\langle f|T_{\pi\gamma}(\vec{q})|i\rangle $ is the amplitude of the
reaction $\pi(\vec {q}) + N \to \gamma + \Delta$ with the free plane wave
of a pion with momentum $\vec{q} \simeq 0$, taken between the initial
$NN$-bound state (i.e. the deuteron) and the final $\Delta N$-state (i.e. the
$d_{1}^{\ast}$- resonance). As we deal with the threshold-type process, we
proceed with keeping only the seagull Feynman graph, approximating the $N\pi
\gamma \Delta$ block ( the corresponding $N\gamma \pi N$-graph is known to
give the low-energy Kroll-Ruderman threshold theorem for the charged pion
photoproduction on nucleons). The $\Delta N\pi$-coupling constant is
defined by the $SU(6)$ symmetry through the known pion-nucleon coupling.
It seems justified also to neglect the retardation corrections,
i.e. we are using the long-wave approximation for the matrix element of
electric-dipole radiative transition.
The $d_{1}^{\ast}\Delta N$-vertex is described by a simple
form of the quasi-two-body wave function, for which the Hulthen-type radial
dependence (1) was chosen by analogy with the deuteron radial wave function.
The radial part of the deuteron wave function is obtained from (1)
when $\varepsilon_1=M+M_{\Delta}-M_{d_{1}^{\ast}}$ is replaced by
$\varepsilon = 2.23$ MeV.
Taking the measured value of the total width $\Gamma_{tot} \simeq 1$ eV \cite
{a11} for the $1S$-level of pionic deuterium, we obtain the following
estimation for the (unobserved) decay channel
\begin{eqnarray}
BR((\pi^{-}p)_{atom} \rightarrow \gamma d_{1}^{\ast} \to \gamma \gamma
~X)_{1S-state} = .6 \%,
\end{eqnarray}
surpisingly close to scale estimate (2) and not
embarrasingly distant from the experimental bound $\le .4 \%$,
despite the adopted approximations being crude.
We note also, that our result does not
depend strongly on variation of the "effective" mass $M_{\Delta}$ from
$M+m_{\pi}$ to $M_{\Delta}=1232 MeV$, ($M=939 MeV, m_{\pi}=139 MeV$ being
masses of the nucleon and pion), if $M_{d_{1}^{\ast}}=1920 $~MeV
is taken for granted.
To get an estimate of the background non-resonance $2\gamma$- emission rate,
we take
\begin{eqnarray}
BR(\pi^{-}d \to 2\gamma/1\gamma) \simeq BR(\pi^{-}p \to 2\gamma/1\gamma)
\simeq 1.3 \times 10^{-4},
\end{eqnarray}

where the corresponding ratio for the pionic hydrogen was calculated
by Beder \cite{a12}.\\
We have then
\begin{eqnarray}
BR((\pi^{-}d)_{atom} \to \gamma \gamma X)_{nonres} \simeq
BR((\pi^{-}d)_{atom} \to \gamma nn) \times 1.3 \cdot 10^{-4}
\simeq 3.4 \cdot 10^{-5}
\end{eqnarray}

which is considerably lower than the estimated resonance contribution.
We point out also a qualitative difference of the $\gamma_1 -
\gamma_2$ opening angle $\theta_{12}$ distribution following from the
resonance and nonresonance mechanisms. In the resonance excitation mechanism,
we have emission of the electric-dipole photon at the $d_{1}^{\ast}$-resonance
excitation vertex and the magnetic-dipole photon emission in the $d_1 \to
\gamma nn$ transition, the nn-pair being mainly in the $^{1}S_{0}$-state. The
polarization structure of the matrix element
\begin{eqnarray}
T(\vec{\epsilon_d}, \vec{\varepsilon_1} (k_1), \vec{\varepsilon_2} (k_2), ...)
\sim \ a_1 ([\vec{\epsilon_d} \times
\vec{\varepsilon_1}] \cdot [\vec{\varepsilon_2} \times \vec{k_2}]) +
(1 \leftrightarrow 2)
\end{eqnarray}

gives after the squaring and summation over polarizations
\begin{eqnarray}
W(\theta_{12},\varphi) = \frac{3}{16 \pi}\cdot (1+ \frac{1}{2} \cdot
\sin^{2}\theta_{12})
\end{eqnarray}
which has a maximum at $\theta_{12}=90^{o}$, while the corresponding
distribution in
the $\pi^{-}p \to 2\gamma n$ reaction, calculated by Beder \cite{a12},
and, by our assumption based on the impulse approximation, also
in the $(\pi^{-}d)_{nonres} \to \gamma \gamma X$ -reaction, shows a shallow
minimum at $\theta_{12}=90^{o}$.\\
It can be noted in this respect that a recent
calculation \cite{a13} of the $\theta_{12}$ -distribution in reactions
$\pi^{-}A \to 2\gamma X$, $(A= ^{9}Be, ^{12}C) $  approximately
agrees with experiment \cite{a14,a15} for angles larger than $90^{o}$ but for
$\theta_{12} \le 90^{o}$ the calculations are consistently lower than data.
A possible role of the exotic resonance excitation is suggestive here, but
for a more quantitative estimation one has to take into account a number of
very important many-body effects: the Pauli blocking, Fermi-motion smearing,
collision broadening of the resonance propagating in nuclear matter. Indeed,
each inelastic $d_{1}^{\ast}N$-collision can trasform the "$\Delta$-part" of
the resonance into a nucleon via isovector, spin-dependent forces trasmitted
by pi- and rho-mesons, thus giving rise to a new decay channel
$d_{1}^{\ast}N \to 3N$.

Qualitatively, instead of $\Gamma^{free}_{tot}(d_{1}^{\ast})=\Gamma_{rad}
(d_{1}^{\ast}) \simeq .5$ KeV we are led to use
$\Gamma^{matter}_{tot}(d_{1}^{\ast}) \simeq \rho \cdot
v \cdot \sigma_{inel}(d_{1}^{\ast}N)$,
and for $\rho \simeq .17 fm^{-3}$, $v  =.2$ and
$\sigma_{inel} \simeq 1$ mb, we get
$\Gamma^{matter}_{tot}(d_{1}^{\ast})$ enhanced by 3
orders of magnitude as compared to $\Gamma^{free}_{tot}$.  That leads to
$BR(\pi^{-}A \to 2\gamma X)$ of the order $\le 10^{-5}$ in accord with
measurements \cite{a14,a15}.  To conclude this section, we are tempted to
mention that the $\gamma$-spectra in radiative pionic deuterium decays can, in
principle, test narrow baryon exotics of the type claimed in ref. \cite{a16},
where the evidence was presented for three narrow baryon states with masses
1004, 1044 and 1094 MeV.  The first two of them are within reach of pionic
mesoatoms studies.

Our conclusion to this section is: In addition to the planned \cite{a17}
measurements of processes $\pi^{-}p \to 2\gamma n $ and double photon
capture on complex nuclei \cite{a18}, the radiative capture processes in
pionic deuterium as well as the continuum pion energy reactions
$\pi^{\pm}d \to \gamma (2\gamma) NN$ well deserve a devoted study
being a perspective source of potentially very important information.\\

\section{Photoexcitation of two-nucleon exotics in nuclei: the local and
integral effects}

The contributions of possible $NN$-decoupled dibaryon resonances to the
photon-deuteron processes like $\gamma d \to \gamma d(\gamma pn)$,
$\gamma d \to \pi^{-}\gamma pp$ have earlier been considered in
ref.\cite{AkhFil}.The emphasis and most of the numerical estimations were made
there for the isoscalar resonances with the quantum numbers
$I=0, J^P=0^{\pm},1^{-}$, while in this section we shall focus on the
$d_{1}^{\ast}(I=1, J^P=1^{+})$-resonance and all illustrative numerical
estimations are based on the specific dynamical model with input
characteristics extracted from available data \cite{a2}.
As far as $\Gamma_{tot} = \Gamma_{rad}$, the value of the
$\gamma d$ elastic scattering cross section is huge at the resonance
value of photon energy $\omega \simeq 45$ MeV, if we take $M_{d_{1}^{\ast}}
= 1920$ MeV.
The theoretical Breit-Wigner cross section with a very small width
will then look like a kind of the "$\delta $-function"-distribution,
while in reality one can observe a much smoother curve
due to the finite initial photon energy spread and finite energy
resolution of a detector. Nevertheless, with the use of extremely high
monochromaticity beams, which can, in principle, be  provided by
the laser devices, the effect of the supernarrow resonance excitation
should be quite spectacular. The off-mass-shell $d_{1}^{\ast}$
contribution to
spin-independent amplitudes of the $\gamma d$ scattering can qualitatively be
estimated through the dynamical (magnetic) polarizability

\begin{eqnarray}
\Delta \beta (\omega) \simeq \frac{\omega_{r}^2 \Delta
\beta_{stat}}{\omega_{r}^2 - \omega^2},\\ \beta_{stat} \simeq \frac{1}{2
\pi^2} \int \frac{d \omega \sigma_{res}^{M1}(\omega)}{\omega ^2} \simeq 2\cdot
10^{-4} fm^{3},
\end{eqnarray}
where the magnetic-dipole $d_{1}^{\ast}$-
photo-excitation cross section, $\sigma_{res}^{M1}$, is approximated by the
Breit-Wigner formula with $\Gamma (d_{1}^{\ast} \to \gamma d) \simeq .2$ KeV,
and $\omega_r = (M_{d_{1}^{\ast}}^2 - M_{d}^2)/2M_{d}$ with $ M_{d_{1}^{\ast}}
\simeq 1920 $ MeV.  For $ \omega < \omega_{r}$ ($ \omega > \omega_{r}$)
$\Delta \beta (\omega)$ is positive (negative), thus contributing to the
paramagnetic (diamagnetic) part of the polarizability coefficient.
Correspondingly, the maximum effect on the differential cross section of the
$\gamma d $- scattering due to $\Delta \beta (\omega) \neq 0$ will then be
seen via its interference with the leading electric-dipole nuclear amplitudes
in photon scattering into the forward (backward) direction. In the interval of
photon energies $50 - 70 $~MeV, the dynamical polarizability due to the exotic
$ d_{1}^{\ast}(1920)$-resonance has the same order of magnitude as
electromagnetic polarizabilities of nucleons. Hence, the interpretation of
experiments on measurements of nucleon polarizabilities from the deuteron
Compton scattering may be influenced considerably by the presence of
low-lying, exotically-narrow six-quark resonance(s).  More detailed
investigation of these questions along the line, {\it e.g.}, of the approach
of ref.\cite {LevLv} is worthwhile in view of experiments planned in Sweden
and Canada ( referred in \cite {LevLv}). In the Compton scattering on heavy
nuclei, the smearing effects due to Fermi-motion of correlated nucleon pairs
and collision broadening of the resonance width have to be taken into account
and they wash out the sharp resonance, but the average enhancement of the
cross section due to possible dibaryon resonance excitation in the
intermediate state is expected to be the same. More careful scanning of the
photon energy interval around $40-50$~MeV in the $\gamma Pb$ - scattering at
large $(\sim 150^{\circ})$ seems to be worthwhile because the Mainz data
\cite{Mz} may give a hint of a weak "shoulder" over smooth background
(Fig.16(d) in ref.\cite {Mz}).

Concerning characteristics of the spin dependence of nuclear photoabsorption,
we mention a possible exotic dibaryon contribution to
the polarized photon - hadron sum rule , which was derived for
hadrons of any spin \cite{GDH} and has started to be applied to nucleons
and lightest nuclei \cite{Ger67} since long time ago ( for a recent
review see \cite {DD} and references therein).
The specific feature of the polarized photon - deuteron sum rule is
the small value of the left-hand side of the sum rule ($ \sim (\mu_{d}
-Q_{d}/M_{d})^2$), where $\mu_{d} (Q_{d})$ is the deuteron magnetic moment
(electric charge),
which should be balanced by very strong compensation of
a large negative value of the integral over deuteron photodisintegration cross
sections below the pion photoproduction threshold \cite {Ger67},\cite{Ar} and
equally large positive value of the rest integral over higher energies
in the right-hand side of the same sum rule. Possible $d_{1}^{\ast}$ -
excitation is immersed into photodisintegration region but it can easily
escape detection due to its extreme narrowness. We just mention here that
with the assumed $d_{1}^{\ast}$ - structure parameters its contribution to the
deuteron sum rule is about $-10~\mu b$, which is of the order of calculated
difference between two above-mentioned and, individually, much larger entries.
To conclude this section: Of indispensable value for direct check of low-lying
exotic dibaryons with isospin $I<2$ there would be scanning of the elastic
$\gamma d$-scattering cross section with tagged photons in the interval of
$40-50$~MeV and with as good energy resolution as possible.

\section{Remarks}

We conclude with the following remarks.\\
1. Among theoretical models predicting dibaryon resonances with different masses
there is one giving the state with the $IJ^P=11^{+}$ and the mass value
$(\sim 1940 MeV)$ surprisingly close to the value $(\simeq 1920 MeV)$
extracted from the observed maximum of the $pp \to pp2\gamma$-reaction.
This is the chiral soliton model applied to the sector with the baryon
number $B=2$ \cite{Kopel95}.  The theoretical uncertainty at
the level of $\pm 30 MeV$ might be taken here because the model gives this
numerical (unrealistic) value for the mass difference of the deuteron
and the singlet level. The absence of this (and some other ) exotic,
$NN$ - decoupled states was claimed to lead to definite restrictions
on the applicability of the chiral soliton approach in the baryon
number $B>1$ - sector. However the cited radiative width of the order
$\sim O(eV)$ \cite{Kopel95} looks much lower compared to our estimate
following from the $N(\pi N)_{33}$ - or $N\Delta_{eff}$ - cluster model \cite
{a6} we have used in this work.

2. Continuation of experiments dealing with closely related processes
such as the deuteron Compton scattering and the double radiative pion-capture
or radiative muon-capture in deuterium mesoatoms would be very helpful not
only as a different area of checking the very existence of exotic dibaryons,
but also as a means to discriminate between possible values of their isospin.

\section*{Acknowledgements}
The author is very indebted to A.M.Baldin, A.S.Khrykin and V.A.Petrun'kin
for discussions of different parts of this report.
This work was supported in part by the Russian Foundation for Basic
Research, grants No. 96-15-96423 and 96-02-19147.

\end{document}